\newif\ifanonymous\anonymousfalse

\documentclass[runningheads]{llncs}
\usepackage[T1]{fontenc}
\usepackage[utf8]{inputenc}
\usepackage{lmodern}
\usepackage{colortbl}
\usepackage{xcolor}
\usepackage{graphicx}
\usepackage{todonotes}
\usepackage{csquotes}
\usepackage[american]{babel}
\usepackage[hidelinks]{hyperref}
\usepackage[acronym]{glossaries}
\usepackage{cleveref}
\usepackage{booktabs}
\usepackage{xspace}
\usepackage{tabularx}
\usepackage{float}
\usepackage{eurosym}
\usepackage[normalem]{ulem}
\usepackage{listings}

\usepackage{xurl}
\usepackage{hyperref}
\hypersetup{breaklinks=true}

\usepackage{color}

\usepackage{tikz}
\usetikzlibrary{
  positioning,
  fit,
  arrows.meta,
  calc
}

\usepackage{caption}

\DeclareCaptionLabelFormat{lncs}{\textbf{#1~#2.}}

\makeatletter
\AtBeginDocument{%
\renewcommand{\ref}[1]{%
  \@latex@error{\string\ref\space nicht verwenden. Bitte durch cleveref (z.B. \string\Cref) ersetzen!}
}}
\makeatother

\newcommand{\eg}{e.g.,\xspace}

\AtBeginDocument{
  \counterwithout{lstlisting}{chapter}%
}
\makeatletter
\renewcommand{\l@lstlisting}[2]{%
  \@dottedtocline{1}{0em}{1.5em}{\lstlistingname\ #1}{#2}%
}
\makeatother

\crefname{lstlisting}{listing}{listings}
\Crefname{lstlisting}{Listing}{Listings}

\lstdefinelanguage{json}{
    basicstyle=\ttfamily\scriptsize,
    numbers=left,
    numberstyle=\tiny,
    stepnumber=1,
    numbersep=8pt,
    showstringspaces=false,
    breaklines=true,
    frame=single,
    backgroundcolor=\color{gray!5},
    escapeinside={(*@}{@*)}
}

\colorlet{altrowcolor}{gray!10}

\begin{document}

\newacronym{dsr}{DSR}{Design Science Research}
\newacronym{adr}{ADR}{Architecture Decision Record}
\newacronym[longplural={Bills of Materials}]{bom}{BOM}{Bill of Materials}
\newacronym[longplural={Cryptographic Bills of Materials}]{cbom}{CBOM}{Cryptographic Bill of Materials}
\newacronym[longplural={Software Bills of Materials}]{sbom}{SBOM}{Software Bill of Materials}
\newacronym{atam}{ATAM}{Architecture Tradeoff Analysis Method}
\newacronym{satam}{SATAM}{Security-Aware Architecture Tradeoff Analysis Method}
\newacronym{sqas}{Security QAS}{Security Quality Attribute Scenarios}
\newacronym{caraf}{CARAF}{Crypto Agility Risk Assessment Framework}

\title{Architecture-Derived \glsentryshortpl{cbom} for Cryptographic Migration: A Security-Aware Architecture Tradeoff Method}

%
\titlerunning{Architecture-Derived \glsentryshortpl{cbom} for Cryptographic Migration}

%
\ifanonymous
  \author{Authors and affiliations withheld for Peer Review\\[1.15cm]}
\else
  \author{%
    Eduard Hirsch\inst{1}\orcidID{0000-0001-9593-2342} \and
    Kristina Raab\inst{2}\orcidID{0009-0001-7072-0340}
  }
  \authorrunning{E. Hirsch et al.}
  \institute{%
    University of Applied Sciences Amberg-Weiden, Amberg 92224, Germany\\
    \email{e.hirsch@oth-aw.de}
    \and
    Fraunhofer AISEC, Garching 85748, Germany\\
    \email{kristina.raab@aisec.fraunhofer.de}
  }
\fi

\maketitle              

\begin{abstract}
Cryptographic migration driven by algorithm deprecation, regulatory change, and post-quantum readiness requires more than an inventory of cryptographic assets. Existing \glspl{cbom} are typically tool- or inventory-derived. They lack architectural intent, rationale, and security context, limiting their usefulness for migration planning. This paper introduces \gls{satam}, a security-aware adaptation of scenario-based architecture evaluation that derives an architecture-grounded, context-sensitive \gls{cbom}. \gls{satam} integrates established approaches: \glsentryshort{atam}, arc42, STRIDE, \glsentryshort{adr}, and \glsentryshort{caraf}. These Generated artifacts are used to annotate \gls{cbom} entries with architectural context, security intent, and migration-critical metadata using CycloneDX-compatible extensions. Following a Design Science Research approach, the paper presents the method design, a conceptual traceability model, and an illustrative application. The results demonstrate that architecture-derived \glspl{cbom} capture migration-relevant context that is typically absent from inventory-based approaches. Thereby, \gls{satam} improves availability of information required for informed cryptographic migration planning and long-term cryptographic agility.
\end{abstract}



\section{Introduction}

Cryptographic systems are subject to continuous change driven by algorithm deprecation, emerging vulnerabilities, regulatory requirements, and the anticipated transition to post-quantum cryptography~\cite{nist_pqc_migration_2025}. As a result, software systems increasingly face the need for systematic cryptographic migration rather than one-time cryptographic upgrades~\cite{naether2024migrating}. However, cryptography is deeply embedded in software architectures, often implicitly and without explicit documentation of its security intent, architectural role, or underlying design rationale. This lack of architectural transparency significantly complicates migration planning and increases the risk of unintended security regressions~\cite{nist_crypto_agility_2019,tyree_akerman_adr_2005}.

Recent efforts to improve cryptographic transparency, such as \glsreset{cbom}\gls{cbom}, aim to enumerate cryptographic algorithms, libraries, and key parameters used within a system. While such inventories are a necessary first step, they are typically derived from tooling or static analysis. They remain largely disconnected from the underlying architecture, following inventory first approaches~\cite{ntia_sbom_framing_2021,OWASP2024CycloneDXCBOMGuide}. As a consequence, existing \glspl{cbom} rarely explain why cryptography is used, which threats components address, or what tradeoffs were considered. Nevertheless, this information is essential for informed decisions~\cite{ntia_sbom_framing_2021}. Without this contextual information, cryptographic migrations are significantly limited, as it becomes difficult to assess the security impact, feasibility, and risk of replacing cryptographic mechanisms. Similarly, established security practices such as threat modeling focus on identifying potential attacks but do not systematically capture architectural decisions or their implications for cryptographic agility.

The \emph{central problem} addressed in this paper is that existing \glspl{cbom} provide limited support for migration-critical architectural questions. In particular, they do not clearly indicate in what sense cryptographic mechanisms are security-relevant or which architectural decisions constrain change. They also fail to show how cryptographic replacements affect system-level security properties and risk assessment.

The \emph{objective of this work} is to design and evaluate a security-aware architecture evaluation method that enables the construction of an architecture-derived \gls{cbom}. During the generation process it is getting enriched with security intent, architectural rationale, and migration-critical metadata like risk assessment. To this end, we introduce \glsreset{satam}\gls{satam}, an approach that adapts scenario-based architecture evaluation to explicitly address cryptographic decisions and their role in mitigating security threats. \gls{satam} integrates threat modeling, security quality attribute scenarios, and architectural decision analysis to make cryptographic assumptions explicit and traceable. The resulting \gls{cbom} is not merely an inventory, but an artifact of the architectural design process that underlies cryptographic migration and long-term agility in complex interdependent systems.

Following a Design Science Research approach, we address the following \emph{design evaluation question}:

\begin{description}
  \item[\textbf{Design Question:}] Does the proposed design artifact \gls{satam} enable the construction of an architecture-grounded, context-sensitive \gls{cbom} by making cryptographic decisions and their security rationale explicit and traceable?
\end{description}

To address this design question, this paper makes the following \emph{contributions}:
(i) \gls{satam}, a security-aware adaptation of scenario-based architecture evaluation that explicitly captures cryptographic decisions and enables the derivation of architecture-grounded, context-sensitive \glspl{cbom};
(ii) a traceability model linking architectural elements, security threats, design decisions, and cryptographic assets; and
(iii) an illustrative integration of \gls{caraf} to demonstrate how architecture-derived artifacts can provide structured input for cryptographic risk assessment.

The method and artifacts are evaluated analytically and through a proof-of-concept, demonstrating how architecture-derived \glspl{cbom} increase transparency and provide structured information to support informed cryptographic migration planning.


\section{Background and Related Work}

\subsection{Architecture Evaluation Method}
\label{sec:background-arch-eval}

Architecture evaluation methods assess whether a software architecture satisfies quality requirements by analyzing architectural structures and design decisions. A prominent example is the \gls{atam}, a scenario-based approach that evaluates stakeholder-defined quality attribute scenarios against architectural approaches to identify risks, tradeoffs, and design rationale~\cite{bass2013atam,Kazman2000ATAM}.

While \gls{atam} is widely applied to performance and modifiability concerns, its treatment of security is typically high-level and does not explicitly address cryptographic mechanisms or migration concerns. Nevertheless, its structured scenario-based reasoning provides a suitable foundation for security-oriented extensions that require architectural context and traceability.

\gls{atam} is independent of specific documentation frameworks but is often combined in practice with arc42~\cite{starke2023arc42}, which structures architectural views, quality requirements, and architectural decisions. In this work, arc42 serves as the baseline architectural description used to identify security-relevant decision points and cryptographic assets.

\subsection{Security Architecture and Threat Modeling}

Security architecture analysis evaluates whether architectural structures and decisions adequately mitigate relevant threats, which are typically defined through explicit threat models. Threat modeling provides a systematic means to identify how architectural elements may be misused or attacked and serves as a key input to security-focused architecture evaluation~\cite{Ibnugraha2015ATAMStride}.

In this work, STRIDE is applied to architectural elements such as components, data flows, and trust boundaries to structure security concerns and ensure systematic threat coverage. STRIDE threats are then transformed into \gls{sqas}~\cite{bass2013atam,Hassouna2025}, which describe required system behavior in response to security-relevant stimuli under defined conditions. \gls{sqas} follow the established scenario structure used in architecture evaluation methods and enable systematic analysis of how architectural decisions and cryptographic mechanisms contribute to security properties.

\subsection{Cryptographic Transparency and Decision Rationale}

Cryptographic transparency has gained increasing attention through the introduction of \glspl{sbom} and, more recently, \glspl{cbom}~\cite{Aroraetal2022StrengtheningOTSecurity,HessKoertge2024CBOMStandardization}. While \glspl{sbom} focus on software components and dependencies, \glspl{cbom} aim to enumerate cryptographic algorithms, libraries, protocols, and key parameters used within a system. Such inventories are valuable for compliance and vulnerability management but are typically derived from tooling or static analysis and remain disconnected from architectural context.

\glspl{adr}~\cite{nygard2011adr} further provide a lightweight and established means to document architectural decisions, including their context, rationale, and consequences. In the context of cryptographic systems, \glspl{adr} can capture the reasoning behind cryptographic choices, such as algorithm selection, key management approaches, or protocol usage. However, \glspl{adr} are typically maintained separately from \glspl{cbom} and are not systematically linked to cryptographic inventories.

\gls{caraf}~\cite{ma2021CARAF} is a proactive approach for analyzing risks arising from limited cryptographic agility, particularly in the context of future threats such as post-quantum cryptography. In this work, \gls{caraf}  is used to illustrate how architecture-derived artifacts can provide structured input for cryptographic risk assessment.



\subsection{Related Work}
\label{sec:related-work}

Prior work on architecture evaluation has demonstrated that scenario-based methods such as \gls{atam} can be extended to incorporate security concerns, for example by integrating threat modeling techniques such as STRIDE~\cite{Ibnugraha2015ATAMStride}. These approaches focus primarily on identifying architectural security risks and tradeoffs rather than producing reusable artifacts for cryptographic transparency or migration planning.

Recent work on Cryptographic Bills of Materials emphasizes the need for systematic cryptographic discovery as a prerequisite for cryptographic migration, particularly in the context of post-quantum cryptography~\cite{BSI2021MigrationPQC,NISTSP1800-38B-PreDraft2023}. \gls{cbom} concepts and data models have been proposed and standardized within initiatives such as OWASP CycloneDX~\cite{OWASP2024CycloneDXCBOMGuide,CycloneDX-CBOM-Capability}. These efforts define the scope of cryptographic inventories and highlight the importance of capturing algorithms, protocols, and key material. Nevertheless, current \gls{cbom} approaches are predominantly tool- or inventory-driven and capture cryptographic usage largely in isolation from architectural context and decision rationale~\cite{HessKoertge2024CBOMStandardization}.

Research has shown that tool-generated \glspl{sbom} often suffer from incompleteness, semantic ambiguity, and limited downstream utility for security decision-making~\cite{Xia2023EmpiricalSBOM,Zhou2025RealityCheckSBOM}. Studies on stakeholder usage indicate that the absence of contextual and semantic information significantly reduces the practical value of \gls{bom} in complex decision-making scenarios~\cite{Stalnaker2024BOMsAway}. Although these studies focus on \glspl{sbom}, their findings are directly applicable to \glspl{cbom} and suggest that inventory-level representations alone are insufficient for supporting cryptographic migration.

Taken together, existing work provides mature methods for scenario-based architecture evaluation, systematic threat identification, and cryptographic inventorying, as well as mechanisms to document architectural decisions and assess cryptographic agility risks. However, these approaches are typically applied in isolation and do not provide a unified method to derive migration-relevant \glspl{cbom} grounded in architectural analysis and security reasoning. This gap motivates the integration of architecture evaluation, threat modeling, decision documentation, and cryptographic inventorying in a single, security-aware evaluation method. \Cref{sec:satam} introduces \gls{satam}, which builds on the concepts outlined in this section and combines them into a coherent process for deriving architecture-grounded, migration-relevant \glspl{cbom}.


\section{Design Science Research Approach}
\label{sec:design-science-research}

This work follows a \gls{dsr} approach~\cite{hevner2004design}, which is well suited for designing and validating artifacts that address practical problems. The primary artifact of this work is \gls{satam}, a security-aware architecture evaluation method for deriving context-sensitive \glspl{cbom}.

Within the \gls{dsr} framework, the problem was identified as the lack of a systematic way to derive migration-relevant \glspl{cbom} from architectural analysis. Existing \gls{cbom} approaches focus on cryptographic discovery but provide limited architectural context and decision rationale, while established architecture evaluation methods offer limited support for cryptographic migration concerns.

Rather than introducing a new inventory format or cryptographic analysis technique, this work focuses on adapting scenario-based architecture evaluation to make cryptographic decisions explicit, analyzable, and traceable. This problem framing directly guided the design of \gls{satam} as a security-aware extension of existing architecture evaluation practice.

\subsection{Artifact Design}

Following \gls{dsr} principles, the artifact was designed by adapting and integrating established concepts from software architecture evaluation and security engineering. \gls{satam} is a method-level artifact that extends scenario-based architecture evaluation with structured threat modeling, security quality attribute scenarios, cryptography-specific analysis, and explicit documentation of architectural decisions.

The design emphasizes traceability between architectural elements, security concerns, cryptographic decisions, and cryptographic assets. This ensures that the resulting \gls{cbom} captures not only which cryptographic mechanisms are used, but also the architectural rationale for their use and the implications for future migration.

In addition, the design allows architecture-derived \gls{cbom} artifacts to provide structured input for a hybrid qualitative and quantitative cryptographic risk assessment, illustrated in this work through the integration of the \gls{caraf}.

\subsection{Evaluation Strategy}

The artifact is evaluated using a combination of analytical and illustrative methods, which is appropriate for an early-stage \gls{dsr} contribution. Analytical evaluation is used to assess internal consistency, traceability, and conceptual completeness of the method, as well as its alignment with established architecture evaluation practices. In addition, an illustrative example is used to demonstrate the applicability of \gls{satam} and to show how architectural analysis results are transformed into an annotated \gls{cbom}.

An empirical evaluation in an industrial setting is outside the scope of this paper and is identified as future work.


\section{\glsreset{satam}\gls{satam} towards \gls{cbom}}
\label{sec:satam}

This section introduces \gls{satam} as a method-level artifact that extends scenario-based architecture evaluation to derive migration-relevant \glspl{cbom}. \gls{satam} builds on \gls{atam} and preserves its overall structure, while introducing explicit security analysis steps and cryptographic decision artifacts. Its objective is not only to identify architectural risks and tradeoffs, but also to produce an annotated \gls{cbom} that captures cryptographic assets together with architectural context, security intent, and migration-relevant metadata.

\begin{table}[!t]
\centering
\caption{\gls{satam} extensions to \gls{atam} and resulting artifacts\\(authoritative process overview)}
\label{tab:satam-overview}
\renewcommand{\arraystretch}{1.3}
\begin{tabularx}{\linewidth}{p{2.1cm} X}
\toprule
\textbf{\gls{atam} step} & \textbf{\gls{satam} activities and key artifacts} \\
\midrule

\multicolumn{2}{X}{\textbf{Preparation phase}} \\
& Identify security/cryptography stakeholders; define security accountability; prepare arc42 architecture description as evaluation baseline\newline
\textbf{Output:} Security roles, \gls{cbom} ownership, security-focused arc42 \\

\midrule
\multicolumn{2}{X}{\textbf{Kickoff phase}} \\
\multicolumn{2}{X}{Presentation of} \\
~~Method
& Position \gls{satam} as \gls{atam} with security annotation; define \gls{cbom} as primary outcome; clarify evaluation scope\newline
\textbf{Output:} Evaluation scope, \gls{cbom} schema \\

~~Objectives
& Treat security explicitly as a business risk; introduce STRIDE as threat assessment input; explain the role of \gls{sqas} and \glspl{adr}\newline
\textbf{Output:} Security drivers, initial \gls{sqas} templates \\

~~Architecture
& Present arc42 architecture in a security-readable form (trust boundaries, crypto termination points, key ownership)\newline
\textbf{Output:} Initial crypto inventory, \gls{adr} candidates \\

\midrule
\multicolumn{2}{X}{\textbf{Evaluation phase}} \\
~~Quality Tree
& Derive the security branch of the quality tree; derive concrete security scenarios by applying STRIDE to architectural elements and flows.\newline
\textbf{Output:} STRIDE-tagged security scenarios \\

~~Analysis
& Refine security scenarios into concrete \gls{sqas}; analyze architectural and cryptographic approaches; apply \gls{caraf} to assess risks based on findings; finalize \glspl{adr}\newline
\textbf{Output:} Final \gls{sqas}, \glspl{adr}, risks, tradeoffs, \gls{caraf} assessment\\

\midrule
\multicolumn{2}{X}{\textbf{Results}} \\
& Architecture baseline, security scenarios, \glspl{adr}, \gls{caraf} findings\newline $\Rightarrow$ integrated into crypto inventory\newline
\textbf{Output:} Annotated, context-sensitive \gls{cbom}\\

\bottomrule
\end{tabularx}
\end{table}

\gls{satam} emphasizes explicit traceability between architectural elements, security concerns, cryptographic decisions, and \gls{cbom} entries. This ensures migration-relevant reasoning across architectural artifacts.

\gls{satam} follows the same high-level phases as \gls{atam} but augments them with security-specific activities and artifacts. In particular, \gls{satam} introduces explicit security stakeholders, treats security as a first-class quality concern, and requires cryptographic mechanisms and key management aspects to be visible in the architectural representation. These extensions ensure that cryptographic assumptions are surfaced early in the evaluation process.

During the evaluation phase, \gls{satam} combines structured threat modeling with scenario-based analysis. STRIDE is applied to architectural elements, data flows, and trust boundaries to identify security-relevant threats, which are then transformed into \gls{sqas}. These scenarios form the basis for analyzing architectural approaches and associated cryptographic mechanisms.

Cryptographic decisions identified during the evaluation are treated as explicit architectural decision candidates and documented using \glspl{adr}. To assess their long-term suitability, \gls{satam} uses \gls{caraf} as a risk analysis lens. \gls{caraf} provides a structured methodology for evaluating cryptographic tradeoffs--specifically agility, transition costs, and quantum-readiness. This process yields a clear profile of identified risks and justified cryptographic decisions for a future-proof security posture.

A key distinction of \gls{satam} is that evaluation results are consolidated into a machine-readable \gls{cbom}. Instead of treating the \gls{cbom} as a standalone inventory, \gls{satam} derives entries directly from architectural analysis artifacts.


\Cref{tab:satam-overview} summarizes the \gls{satam} steps and highlights the extensions beyond standard \gls{atam}. \Cref{tab:satam-caraf-mapping} shows how all \gls{caraf} inputs, except adversarial capability assumptions, are systematically derived from \gls{satam} artifacts.

\begin{table}[h]
\centering
\scriptsize
\caption{Mapping of \gls{satam} artifacts to \gls{caraf} inputs\newline
             for cryptographic migration risk assessment}
\label{tab:satam-caraf-mapping}
\begin{tabularx}{\linewidth}{>{\bfseries}>{\raggedright\arraybackslash}p{2cm}>{\raggedright\arraybackslash}p{3cm} X}
\toprule
\textbf{\gls{caraf} input} & \textbf{Provided by\newline \gls{satam} artifact} & \textbf{Explanation} \\
\midrule
Cryptographic asset &
Architecture-derived \gls{cbom} entry &
Cryptographic mechanisms derived from security-readable architecture; consolidated as \gls{cbom} entries for risk analysis. \\

\rowcolor{altrowcolor}
Protected\newline asset &
arc42 architecture views; \gls{cbom} entry &
Architectural elements and data flows; concrete system context of protected assets. \\

Threat\newline rationale &
STRIDE-tagged scenarios &
Threat categories (e.g., tampering, disclosure, privilege escalation); motivation for cryptographic protection. \\

\rowcolor{altrowcolor}
Security\newline requirements &
\glsreset{sqas}\gls{sqas} &
Measurable criteria (e.g., TLS~1.3 enforcement, downgrade resistance); long-term suitability constraints. \\

Cryptographic\newline decision &
\glsreset{adr}\glspl{adr} &
Selected cryptographic approach; rationale; architectural consequences. \\

\rowcolor{altrowcolor}
Migration\newline time\newline drivers ($Y$) &
\gls{adr} consequences; deployment and operational notes &
Rollout constraints; certificate life cycles; interoperability dependencies; operational coupling. \\

Data\newline lifetime ($X$) &
\gls{sqas} assumptions; regulatory context &
Required confidentiality and integrity durations; long-term protection needs. \\

\rowcolor{altrowcolor}
Cryptographic\newline horizon ($Z$) &
External assumptions (e.g., PQC timelines) &
Adversarial capability horizon; contextualized by \gls{cbom} assets and requirements. \\

Risk\newline mitigation\newline options &
\gls{caraf} findings linked to \glspl{adr} &
Actionable mitigation directions; direct reference to architectural decisions. \\
\bottomrule
\end{tabularx}
\end{table}

\paragraph{Essential Elements of \gls{satam}.}

\gls{satam} integrates established techniques such as arc42, STRIDE, \gls{sqas}, \glspl{adr}, and \gls{caraf}. However, not all of these are methodologically mandatory. The essential contribution of \gls{satam} lies in four minimal requirements:

\begin{enumerate}
    \item A security-readable architecture baseline in which cryptographic assets can be located in architectural context.
    \item Explicit identification of cryptographic assets and their protected elements or data flows.
    \item A traceability chain linking threats or security scenarios to architectural elements, cryptographic mechanisms, and architectural decisions.
    \item A machine-readable \gls{cbom} representation that preserves this traceability.
\end{enumerate}

Specific techniques (e.g., STRIDE for threat modeling, arc42 as documentation template, or \gls{caraf} for risk estimation) are replaceable modules. The novelty of \gls{satam} lies in consolidating architecture evaluation outputs into a traceability-preserving, migration-relevant \gls{cbom} rather than in the individual techniques themselves.


\section{Conceptual Traceability Model for Architecture-Derived \glspl{cbom}}
\label{sec:traceability-model}

This section defines the conceptual traceability model that specifies the structure of architecture-derived \glspl{cbom} produced by \gls{satam}. The model does not introduce additional analysis steps. Instead, it defines the mandatory relationships between architectural artifacts and \gls{cbom} entries required to ensure migration relevance.

The \emph{traceability model defines} the following core relationships:
(i) cryptographic assets are derived from and bound to architectural elements;
(ii) security threats identified using STRIDE are associated with architectural elements and data flows;
(iii) identified threats are refined into \gls{sqas};
(iv) cryptographic design decisions are documented as \glspl{adr} in response to \gls{sqas};
(v) \glspl{adr} are evaluated using \gls{caraf} to assess migration-relevant risk; and
(vi) \gls{cbom} entries reference cryptographic assets together with their associated architectural elements, \gls{sqas}, \glspl{adr}, and \gls{caraf} assessments.

The architecture description (arc42, see \Cref{sec:background-arch-eval}) forms the structural foundation of the traceability model. Architectural elements, including components, data flows, deployment nodes, and trust boundaries, act as anchors for all subsequent traceability links.

Security threats are collected by applying STRIDE to architectural elements and interactions. STRIDE threats represent potential violations of security properties but do not yet specify required system behavior. To enable architectural evaluation, identified threats are transformed into \gls{sqas}. \gls{sqas} describe the expected system response to security-relevant stimuli under defined conditions and provide measurable criteria for assessing architectural suitability.

Architectural responses to \gls{sqas} often involve explicit cryptographic choices, such as algorithm selection, protocol usage, or key management strategies. These choices are captured as \glspl{adr}, which document decision context, rationale, and consequences. Within \gls{satam}, \glspl{adr} act as the primary carriers of cryptographic design intent and form a critical link between security requirements and concrete architectural solutions.

Within the traceability model, \gls{caraf} provides qualitative and quantitative assessments of cryptographic decisions documented in \glspl{adr}. These assessments capture migration-relevant risk considerations and constraints that influence the feasibility and timing of cryptographic change.

The \gls{cbom} is derived by consolidating these traceability relationships. The \gls{cbom} entry corresponds to a cryptographic asset identified in the architecture. Each is annotated with references to the architectural elements it protects and the \gls{sqas} it supports. Additionally, it includes the STRIDE threats it mitigates, the \glspl{adr} justifying its selection, and the \gls{caraf} findings that characterize its migration impact. Through this integration, the \gls{cbom} becomes an architectural artifact that reflects both the technical composition and the security rationale of cryptographic usage.

By making these relationships explicit, the traceability model enables validation of \gls{cbom} completeness and consistency. The resulting \gls{cbom} is not a static inventory, but an architectural artifact reflecting security intent and migration constraints.

\paragraph{Maintenance and Lifecycle.}
\gls{satam} is designed as a continuous process rather than a one-time evaluation activity. All artifacts are anchored in the architecture baseline and evolve together with it. Changes in architectural elements propagate along traceability links and identify affected \gls{sqas}, \glspl{adr}, and cryptographic assets. The \gls{cbom} is treated as a derived artifact and is regenerated from the underlying model. This approach reduces manual maintenance effort and ensures consistency by construction.

\paragraph{Re-evaluation Triggers.}
\gls{satam} supports incremental re-evaluation based on defined triggers. \emph{Architecture-driven triggers} include changes in components, interfaces, data flows, or trust boundaries. \emph{Threat-driven triggers} arise from updated threat models or new attack vectors. \emph{Cryptography-driven triggers} include algorithm deprecation and regulatory changes. \emph{Operational triggers} include changes in deployment, certificate lifecycle, or key management. These triggers enable targeted updates while preserving overall consistency.


\section{Proof of Concept}
\label{sec:proof-of-concept}


\subsection{Example arc42 Architecture Baseline}

\vspace*{-5mm}
\begin{figure}[h]
\centering
  \resizebox{\width}{!}{\pgfdeclarelayer{background}
\pgfsetlayers{background,main}

\definecolor{boundary-color}{gray}{0.3}

\begin{tikzpicture}[
  node distance=15mm and 15mm,
  box/.style={draw, rectangle, rounded corners, minimum width=16mm, minimum height=9mm, align=center},
  boxsmall/.style={box, minimum width=12mm, minimum height=6mm, node distance=5mm and 15mm},
  line/.style={latex-latex, thick},
  directed/.style={-latex, thick}
]
\node[box] (client) {External\\Client};
\node[box, right=of client] (api) {API\\Gateway};
\node[box, right=of api, fill=white] (svc) {OT-Integration\\Service};

\begin{pgfonlayer}{background}
  \draw[dashed, draw=boundary-color]
    ($(client.north)+ (1.05,6mm)$) -- ++(0,-4.2)
    node[below, right, rotate=90, yshift=-2mm, text=boundary-color] {trust boundary $0$};
\draw[dashed, draw=boundary-color]
    ($(api.north)   + (1.05,6mm)$) -- ++(0,-4.2)
    node[below, right, rotate=90, yshift=-2mm, text=boundary-color] {trust boundary $1$};
\draw[dashed, draw=boundary-color]
    ($(svc.north)   + (1.50,6mm)$) -- ++(0,-4.2)
    node[below, right, rotate=90, yshift=2mm, text=boundary-color] {trust boundary $2$};
\end{pgfonlayer}

\node[boxsmall, right=of svc, yshift=7mm] (plc0) {Asset $0$};
\node[boxsmall, below=of plc0] (plc1) {Asset $1$};
\node[boxsmall, below=of plc1] (dots) {$\hdots$};
\node[boxsmall, below=of dots] (plcn) {Asset $n$};

\node[box, below=of api] (idp) {Identity\\Provider};

\draw[line] (client) -- node[above, xshift=1mm]{HTTPS} (api);
\draw[line] (api) -- node[above, align=center, yshift=-4mm]{TLS\\SASL} (svc);

\draw[directed] (svc) -- node[above]{} (plc0);
\draw[directed] (svc) -- node[above]{} (plc1);
\draw[directed] (svc) -- node[above, rotate=-90, xshift=-4mm, yshift=0mm, align=center]{\small OPC\\UA/DA} (plcn);

\draw[line] (api) -- node[sloped, below]{OIDC} (idp);
\draw[line, dashed] (client) to[out=-50, in=180] (idp);
\end{tikzpicture}}%

\caption{Security-readable architecture baseline used as \gls{satam} input}
\label{fig:ex-arch}
\end{figure}

We consider a simplified industrial architecture where an external client interacts with an OT environment via an API gateway and an OT integration service. The integration service communicates with industrial assets using OPC~UA. This design follows established OPC~UA-based industrial data architectures \cite{Hirsch2023industrialBigData}, where OPC~UA is used for both control interaction and data acquisition across heterogeneous devices.

\Cref{fig:ex-arch} shows the security-readable architecture baseline. It makes trust boundaries and cryptographic termination points explicit. This view corresponds to a building block view based on an arc42 documentation~\cite{starkeHruschka2024arc42,starke2023arc42}.

\subsection{STRIDE Threat Collection and how it Drives \gls{sqas}}

\gls{satam} applies STRIDE to architectural elements and flows. This yields threats that are specific to the architecture, not generic checklists. \Cref{fig:ex-stride} illustrates threat collection on two flows: the public client-to-gateway (edge) and internal service-to-assets flow (svc-assets).

\begin{figure}[h]
\centering
  \resizebox{\width}{!}{\begin{tikzpicture}[
  node distance=14mm and 15mm,
  box/.style={draw, rectangle, rounded corners, minimum width=19mm, minimum height=9mm, align=center},
  tag/.style={draw, rounded corners, inner sep=2pt, align=left},
  line/.style={latex-latex, thick}
]
\node[box] (client) {External\\Client};
\node[box, right=of client] (api) {API\\Gateway};
\node[box, right=of api, xshift=-10mm] (svc) {OT-Integration\\Service};
\node[box, right=of svc] (opcserver) {Assets};

\draw[line] (client) -- node[above]{HTTPS} (api);
\draw[line] (svc) -- node[above]{OPC UA} (opcserver);

\node[tag, below=of $(client)!0.5!(api)$] (t1) {\small \textbf{STRIDE (edge)}\\
S: token replay/spoofed client\\
T: request tampering\\
R: insufficient audit evidence\\
D: rate-limit bypass,\\
~~~~request flooding};
\node[tag, below=of $(svc)!0.5!(opcserver)$] (t2) {\small \textbf{STRIDE (svc-assets)}\\
S: OPC UA peer spoofing\\
T: command/telemetry manipulation\\
I: ~disclosure of industrial process data,\\
~~~~esp. if unencrypted (OPC DA)\\
D: endpoint unavailability due to\\
~~~~certificate/session failure\\
E: unauthorized control action via\\
~~~~excessive privileges};

\draw[-, thin, dashed] (t1.north) -- (t1 |- api);
\draw[-, thin, dashed] (t2.north) -- (t2 |- svc);
\end{tikzpicture}}%

\caption{STRIDE threats identified in architecture flows}
\label{fig:ex-stride}
\end{figure}

STRIDE threats are then converted into \gls{sqas}. This conversion makes the analysis actionable, because \gls{sqas} specify required system behavior and measurable acceptance criteria. For example, the STRIDE \textit{tampering} threat on the service-to-assets flow yields a \gls{sqas} primarily addressing integrity while also capturing timing and availability constraints (see~\Cref{tab:QAS-OT-OPC-1}).

\subsection{\gls{sqas} and how it Feeds \gls{caraf}}

\Cref{tab:QAS-OT-OPC-1} shows the \gls{sqas} derived from the STRIDE \textit{tampering} threat (\Cref{fig:ex-stride}) identified for the service-to-assets flow. The response measures (RMs) are intentionally concrete, because they become an evaluation criteria during the \gls{caraf}-based analysis.

\begin{table}[h]
\caption{\gls{sqas} derived from STRIDE threats}
\label{tab:QAS-OT-OPC-1}
\centering
\begin{tabularx}{\linewidth}{>{\bfseries}lX}
\toprule
\multicolumn{2}{l}{\textbf{QAS-OT-OPC-1: Secure and timely industrial communication}} \\
\midrule
Source & Attacker in IT or OT network \\
\rowcolor{altrowcolor}
Stimulus & Injection, modification, or eavesdropping of OPC~UA messages or unprotected OPC~DA data (using \texttt{Sign} only)\\
Environment & OPC~UA communication between OT Integration Service and PLC under normal operation and certificate rotation \\
\rowcolor{altrowcolor}
Artifact & OPC~UA and legacy OPC~DA communication channels \\
Response & Authenticated and unmodified messages are enforced for secure channels, while risks from legacy communication are controlled to maintain stable operation \\
\rowcolor{altrowcolor}
RMs & Auth failures rejected at handshake; integrity violations detected within one cycle; latency $\leq X$ ms; availability $\geq 99.9\%$ during certificate lifecycle events; OPC~DA confined to high-risk trusted zones \\

\bottomrule
\end{tabularx}
\end{table}

This \gls{sqas} drives the \gls{caraf} evaluation by defining measurable criteria for secure and timely OPC communication. It constrains architectural options across latency, availability, and compatibility dimensions but is also taking legacy constraints into account. In our example, enforcing authentication and integrity under timing constraints restricts the choice of OPC~UA security modes. Further, it requires compensating controls for legacy communication, impacting cryptographic agility and operational effort.

\subsection{\glspl{adr} and \gls{caraf} Evaluation}

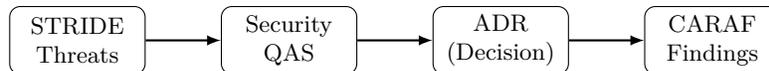
\begin{figure}[h]
\centering
  \resizebox{\width}{!}{\begin{tikzpicture}[
  node distance=5mm and 10mm,
  box/.style={draw, rectangle, rounded corners, minimum width=18mm, minimum height=9mm, align=center},
  line/.style={-latex, thick}
]
\node[box] (stride) {STRIDE\\Threats};
\node[box, right=of stride] (qas) {Security\\QAS};
\node[box, right=of qas] (adr) {\gls{adr}\\Decisions};
\node[box, right=of adr] (caraf) {\gls{caraf}\\Findings};

\draw[line] (stride) -- (qas);
\draw[line] (qas) -- (adr);
\draw[line] (adr) -- (caraf);

\node[below=of qas, align=left] (note) {\small Example chain:\\
\texttt{T/I/S/E} on svc--assets $\rightarrow$ \texttt{QAS-OT-OPC-1} $\rightarrow$ \texttt{ADR-OT-OPC-1}\\
$\Rightarrow$ CARAF-OT-OPC-1 (action required, Risk \EUR{600{,}000})};
\end{tikzpicture}}%

\caption{\gls{satam} evaluation chain}
\label{fig:ex-eval}
\end{figure}

\Cref{fig:ex-eval} makes the \gls{satam} evaluation chain explicit. STRIDE contributes the threat rationale, \gls{sqas} provide measurable requirements, \glspl{adr} capture the chosen architectural response, and \gls{caraf} records migration-relevant findings.

\paragraph{\gls{adr} example.}
\texttt{ADR-OT-OPC-1} (in \Cref{tab:adr-ot-opc-1}) documents the decision to enforce OPC~UA \texttt{SignAndEncrypt} where supported and to isolate legacy OPC~DA within trusted zones. It explicitly references the STRIDE threats (\texttt{T,I,S,E}) and the \gls{sqas} \texttt{QAS-OT-OPC-1} as justification. The decision reflects a tradeoff between confidentiality, integrity, and authentication on the one hand, and latency, availability, and legacy compatibility on the other.

\paragraph{\gls{caraf} example.}
\Cref{tab:caraf-ot-opc-1} shows an excerpt of \gls{caraf} findings derived from \texttt{ADR-OT-OPC-1} in the light of \texttt{QAS-OT-OPC-1} and the STRIDE rationale. The example focuses on migration-relevant aspects, including the effort of adopting secure OPC~UA modes and handling legacy OPC~DA. It is important to note that this excerpt is illustrative and demonstrates how architecture-derived artifacts populate the risk model, rather than providing a calibrated risk estimate.


\begin{table}[h]
\caption{Condensed architecture decision record}
\label{tab:adr-ot-opc-1}
\centering
\begin{tabularx}{\linewidth}{>{\bfseries\arraybackslash}lX}
\toprule
\multicolumn{2}{l}{\textbf{ADR-OT-OPC-1: OPC communication security}} \\
\midrule
Status & Accepted \\
\rowcolor{altrowcolor}
Context & OT-integration-to-PLC communication crosses trust boundary; legacy OPC~DA included; STRIDE identified TISE-threats; QAS-OT-OPC-1: integrity and authentication under latency and availability constraints\\
Decision & Use OPC~UA \texttt{SignAndEncrypt} for service-to-assets where supported; isolate/restrict OPC~DA within trusted zones. \\
\rowcolor{altrowcolor}
Rationale & Prioritizes confidentiality, integrity, and authentication (QAS-OT-OPC-1), trading higher latency and complexity for reduced exposure of process data and control commands; legacy and timing constraints handled via selective deployment and compensating controls. \\
Consequences & Confidentiality, integrity, and authentication for OPC~UA where supported; increased latency and operational complexity; certificate management required; OPC~DA remains unprotected and requires isolation and monitoring. \\
\rowcolor{altrowcolor}
References & \textbf{STRIDE}: TISE(D); \textbf{QAS}: QAS-OT-OPC-1; \textbf{CARAF}: CARAF-OT-OPC-1 \\
\bottomrule
\end{tabularx}
\end{table}

\begin{table}[h]
\centering
\caption{Example risk record based on Mosca's Theorem used within \glsentryshort{caraf}}
\label{tab:caraf-ot-opc-1}

\begin{tabularx}{\linewidth}{>{\bfseries\arraybackslash}p{1.9cm}X}
\toprule
\multicolumn{2}{l}{\textbf{CARAF-OT-OPC-1: \gls{caraf} risk record}} \\
\midrule

Protected asset & Communication channel svc-assets and industrial process data (QAS-OT-OPC-1) \\
\rowcolor{altrowcolor}
Mosca\newline parameters &
Determine Mosca parameters ($X, Y, Z$) based on data lifetime, migration time for OPC~UA security modes and OPC~DA replacement, and quantum horizon \\

Risk check &
Exposure of sensitive process data and control commands,\newline
assumption of $X + Y > Z$ $\Rightarrow$ \textbf{action required} \\

\rowcolor{altrowcolor}
Drivers\newline of $Y$ & Derived from ADR-OT-OPC-1 consequences and operational constraints: certificate management for OPC~UA, legacy OPC~DA client replacement, compatibility constraints, phased rollout across OT assets \\

Risk~\cite{ma2021CARAF} &
$\text{Risk} = T_{\text{r}} \times \text{Cost}$, cost model \eg based on~\cite{Algarni2021Quantitative,IBM2025Cost}\\

\rowcolor{altrowcolor}
Mitigation \newline
direction &
Reduce $Y$ via crypto agility (standardized OPC~UA security policies, automated certificate handling, asset inventory); plan migration from OPC~DA to secure OPC~UA; evaluate lightweight or phased encryption adoption \\

References &
Assessment based on\newline \textbf{STRIDE}: \texttt{TISE(D)}; \textbf{QAS}: QAS-OT-OPC-1; \textbf{ADR}: ADR-OT-OPC-1 \\
\bottomrule
\end{tabularx}
\end{table}

\subsection{Deriving the Annotated \glsentryshort{cbom}}
\label{sec:cbom-derivation}

The \gls{satam} evaluation chain produces a \gls{cbom} entry for each cryptographic asset annotated with stable references. The information can now be consolidated into a structured \gls{cbom}.




\Cref{lst:cbom-json} summarizes the identified cryptographic assets together with their architectural context and traceability links to STRIDE threats, \gls{sqas}, \glspl{adr} as well as \gls{caraf} findings. These \gls{caraf}-derived enrichments are considerations  such as agility, future-proofing, and key risks. Together, these represent a logical decomposition of the assessed information and are consolidated into a single \gls{cbom} artifact.

The JSON in \Cref{lst:cbom-json} shows a simplified JSON example encoded according to the CycloneDX~1.7 specification. Each cryptographic asset is represented as a component and annotated using component properties to reference \gls{satam}-specific metadata. These properties include architectural context, STRIDE threat categories, references to \gls{sqas}, \glspl{adr}, and \gls{caraf} findings related to migration impact.

That representation preserves the traceability relationships shown in \Cref{lst:cbom-json} using CycloneDX-compatible properties. For example, the OPC~UA \texttt{SignAndEncrypt} entry references the same identifiers as declared during the \gls{satam} evaluation. These are identifiers applied for STRIDE threats, \gls{sqas}, \gls{adr}, but also for \gls{caraf} notes on risk assessment. The corresponding OPC~DA entry captures the absence of cryptographic protection and associated risks within the same framework. This ensures consistency between human-readable and machine-readable \gls{cbom} representations.

\gls{satam} thus produces an artifact that supports both expert review and automated migration analysis. The resulting \gls{cbom} bridges architecture evaluation and cryptographic migration tooling while preserving architectural rationale. This proof of concept demonstrates that \gls{satam} can be applied end-to-end to a realistic architecture, yielding an architecture-grounded \gls{cbom} with explicit traceability and migration-relevant annotations.


\section{Evaluation}

\gls{satam} is evaluated analytically and through a proof-of-concept demonstration. As a methodological contribution, the evaluation focuses on internal consistency, traceability completeness, and feasibility rather than industrial-scale empirical validation.

\subsection{Analytical Evaluation}

The analytical evaluation assesses whether \gls{satam} fulfills its intended objectives and addresses the identified design question. First, \gls{satam} is internally consistent with established scenario-based architecture evaluation methods. It preserves the core \gls{atam} phases while introducing security-specific extensions that are methodologically coherent and grounded in established practices (see \Cref{sec:satam,sec:traceability-model}).

\begin{lstlisting}[language=json,
                    caption={Minimal \gls{cbom} JSON excerpt of components\\highlighting the OPC~UA versus OPC~DA tradeoff},
                    label={lst:cbom-json},]
{
  "bom-ref": "CBOM-OT-OPCUA-1", "type": "cryptographic-asset",
  "name": "OPC~UA Security", "version": "SignAndEncrypt", (*@\underbar{\textbf{"properties":}}@*) [
    { "name": "satam.context.flow", "value": "svc-assets" },
    { "name": "satam.stride", "value": "S,T,I,D,E" },
    { "name": "satam.securityQasRefs", "value": "QAS-OT-OPC-1" },
    { "name": "satam.adrRefs", "value": "ADR-OT-OPC-1" },
    { "name": "satam.caraf.id", "value": "CARAF-OT-OPC-1" },
    { "name": "satam.caraf.risk", "value": "150000 Euro" },
    { "name": "satam.caraf.risk.detail",
      "value": "increased latency due to encryption" }
  ]
}, {
  "bom-ref": "CBOM-OT-OPCDA-1", "type": "cryptographic-asset",
  "name": "OPC~DA", "version": "legacy", (*@\underbar{\textbf{"properties":}}@*) [
    { "name": "satam.context.flow", "value": "svc-assets" },
    { "name": "satam.stride", "value": "S,T,I,E" },
    { "name": "satam.securityQasRefs", "value": "QAS-OT-OPC-1" },
    { "name": "satam.adrRefs", "value": "ADR-OT-OPC-1" },
    { "name": "satam.caraf.id", "value": "CARAF-OT-OPC-1" },
    { "name": "satam.caraf.risk", "value": "500000 Euro" },
    { "name": "satam.caraf.risk.detail",
      "value": "eavesdropping and manipulation of process data" }
  ]
}
\end{lstlisting}


Second, \gls{satam} provides explicit traceability across architectural elements, security concerns, cryptographic decisions, and \gls{cbom} entries, as defined \Cref{sec:traceability-model}. This traceability directly addresses the limitations of inventory-based \glspl{cbom} and supports reasoning about cryptographic migration impact. In contrast to inventory-first \glspl{cbom}, \gls{satam}-derived \glspl{cbom} do not merely enumerate cryptographic mechanisms, but make explicit why cryptography is used, which security properties it enforces, and how architectural decisions constrain change.

Third, the method is suitable for its intended purpose of supporting cryptographic migration. \gls{satam} explicitly captures decision rationale and migration-relevant tradeoffs. This enables \gls{satam} architectural reasoning about change impact, dependency constraints, and algorithm agility, as illustrated by the \gls{caraf}-based migration analysis in \Cref{sec:proof-of-concept}. As a result, \gls{satam}-derived \glspl{cbom} expose migration-critical constraints, such as dependency coupling, rollout effort, and key management life cycles, which are typically opaque in tool-generated inventories.

\subsection{Illustrative Evaluation}

The proof of concept in \Cref{sec:proof-of-concept} demonstrates the applicability of \gls{satam} on a realistic, yet manageable, system. The example shows the process of deriving STRIDE threats from the architecture and transforming them into \gls{sqas}. It further illustrates how cryptographic decisions are documented as \glspl{adr} and how \gls{caraf} is applied to assess migration-relevant properties. The resulting \gls{cbom} excerpt and JSON representation confirm that the method produces a coherent, machine-readable artifact that preserves architectural rationale.

While the example does not provide empirical evidence of effectiveness in practice, it demonstrates feasibility and clarity of the approach. This level of evaluation is appropriate for an early-stage \gls{dsr} contribution and provides a foundation for future empirical studies.

\paragraph{Limitations and Validity.} The evaluation is limited to a proof-of-concept instantiation on a compact reference architecture and relies on analytical reasoning rather than empirical measurement. The effort required to apply \gls{satam} and the scalability of the approach for large, evolving systems have not been empirically assessed. Additionally, \gls{caraf} inputs and cost parameters in the example are illustrative and context-dependent. Future work includes empirical validation in industrial settings and longitudinal studies of cryptographic migration decisions. \gls{satam}  partially supports automation, as \gls{cbom} generation, traceability linking, and consistency checks can be derived from the underlying architectural artifacts. Security scenario definition and architectural decision-making remain expert-driven.

\subsection{Task-Based Migration Comparison}

To illustrate practical impact, we compare an inventory-derived \gls{cbom} and a \gls{satam}-derived \gls{cbom} using representative migration tasks.

\begin{table}[h]
\centering
\caption{Task-based comparison of migration support in the OT example}
\footnotesize
\begin{tabularx}{\textwidth}{>{\raggedright\bfseries}p{2.3cm}>{\raggedright}p{3.9cm} X}
\toprule
\textbf{Migration\\ task} & \textbf{Inventory-derived \gls{cbom}} & \textbf{\gls{satam}-derived\newline \gls{cbom}} \\
\midrule
Replace legacy OPC~DA with secure OPC~UA &
Identifies protocol presence only. Trust boundaries, affected flows, and security consequences must be reconstructed manually. &
Identifies the affected OT flow, linked STRIDE threats, referenced \gls{sqas}, and \gls{adr} rationale. Migration scope, residual exposure, and required compensating controls get visible. \\

\rowcolor{altrowcolor}
Strengthen OPC~UA from \texttt{Sign} to \texttt{SignAndEncrypt} &
Lists OPC~UA usage only. Performance, compatibility, and deployment implications assessed externally. &
Links the change to the relevant OT communication flow, \gls{sqas} constraints, and \gls{adr} tradeoff. Migration impact on latency, certificate handling, and rollout complexity is directly visible. \\

OT~security prioritization (availability-constrained) &
Requires separate risk estimation. Legacy exposure and migration effort are not traceable from the inventory alone. &
Provides traceability to security scenarios and \gls{caraf}-based migration risks, supporting prioritization that accounts for security benefit, operational effort, and legacy constraints. \\
\rowcolor{altrowcolor}
Upgrade HTTPS to PQC-safe TLS &
Identifies HTTPS usage only. Affected endpoints and migration dependencies must be reconstructed manually. &
Identifies the client-to-gateway flow, linked STRIDE threats, and \gls{sqas}/\gls{adr} context. Supports assessment of PQC readiness, rollout scope, and impact on interoperability. \\
\bottomrule
\end{tabularx}
\end{table}

The comparison does not claim quantitative performance improvement. It illustrates that \gls{satam} reduces information reconstruction effort by preserving architectural intent and security rationale within the \gls{cbom} artifact itself.


\section{Conclusion}

This paper presented the Security-Aware Architecture Tradeoff Analysis Method for deriving architecture-grounded, migration-relevant \glspl{cbom}. In contrast to existing inventory-driven approaches, \gls{satam} treats cryptographic transparency as an architectural concern and derives \gls{cbom} entries directly from scenario-based architecture evaluation. \gls{satam}  links cryptographic assets to architectural elements, security threats, security quality attribute scenarios, and architectural decisions. This results into \glspl{cbom} embedding security rationale and design intent that are essential for informed cryptographic migration.

The central contribution of this work is the consolidation of architectural evaluation outcomes into a structured \gls{cbom} artifact. Unlike tool- or inventory-derived \glspl{cbom}, the architecture-derived \gls{cbom} produced by \gls{satam} makes cryptographic usage understandable in its architectural and security context and exposes constraints that influence the feasibility of change. This positions the \gls{cbom} not only as a compliance or discovery artifact, but as a practical input for cryptographic migration planning and long-term cryptographic agility.

\paragraph{Limitations and Future Work.}
The evaluation is limited to an analytical assessment and a proof-of-concept instantiation, and no industrial case study has yet been conducted. As a result, the practical effort, scalability, and cost-benefit characteristics of \gls{satam} in real-world environments remain to be validated. In addition, the current presentation assumes familiarity with
architecture documentation and evaluation practices, which may limit applicability in organizations without established architectural governance.



Future research will empirically evaluate \gls{satam} in industrial contexts through comparative studies with automated \gls{cbom} generation tools. Key priorities include developing tool support to automate the \gls{satam} workflow, refining \gls{cbom} profiles for emerging requirements such as post-quantum cryptography, and integrating the approach into secure development pipelines for continuous architecture evaluation. While \gls{satam} introduces upfront analysis effort, it can reduce repeated manual effort by preserving architectural context and enabling partial automation of artifact generation and consistency checking.

The results indicate that, \gls{satam} enables the construction of architecture-grounded, migration-relevant \glspl{cbom} by making cryptographic decisions and their security rationale explicit and traceable.

\bibliographystyle{splncs04}
\bibliography{misc/arch-review-refs}

\end{document}